# AI-Driven Personalised Offloading Device Prescriptions: A Cutting-Edge Approach to Preventing Diabetes-Related Plantar Forefoot Ulcers and Complications

**Sayed Ahmed**, Faculty of Health, Southern Cross University, Billinga, QLD, Australia
Foot Balance Technology Pty Ltd, Westmead, NSW, Australia
**Muhammad Ashad Kabir**, School of Computing, Mathematics and Engineering, Charles Sturt University, Bathurst, NSW, 2795, Australia (akabir@csu.edu.au)
**Muhammad E. H. Chowdhury,** Department of Electrical Engineering, Qatar University, Doha, 2713, Qatar (mchowdhury@qu.edu.qa)
**Susan Nancarrow**, Faculty of Health, Southern Cross University, Billinga, QLD, Australia (Susan.Nancarrow@scu.edu.au)

Corresponding author's mail: s.ahmed.13@student.scu.edu.au, sayed@footbalancetech.com.au

**Abstract**

Diabetes-related foot ulcers and complications are a significant concern for individuals with diabetes, leading to severe health implications such as lower-limb amputation and reduced quality of life. This chapter discusses applying AI-driven personalised offloading device prescriptions as an advanced solution for preventing such conditions. By harnessing the capabilities of artificial intelligence, this cutting-edge approach enables the prescription of offloading devices tailored to each patient's specific requirements. This includes the patient's preferences on offloading devices such as footwear and foot orthotics and their adaptations that suit the patient's intention of use and lifestyle. Through a series of studies, real-world data analysis and machine learning algorithms, high-risk areas can be identified, facilitating the recommendation of precise offloading strategies, including custom orthotic insoles, shoe adaptations, or specialised footwear. By including patient-specific factors to promote adherence, proactively addressing pressure points and promoting optimal foot mechanics, these personalised offloading devices have the potential to minimise the occurrence of foot ulcers and associated complications. This chapter proposes an AI-powered Clinical Decision Support System (CDSS) to recommend personalised prescriptions of offloading devices (footwear and insoles) for patients with diabetes who are at risk of foot complications. This innovative approach signifies a transformative leap in diabetic foot care, offering promising opportunities for preventive healthcare interventions.

Definition

Pedorthist: A person who provides medical-grade footwear and/or orthotic appliances and appropriate advice to a patient after assessment and analysis of the patient's problem(s). This includes the provision of prefabricated footwear, modification of prefabricated footwear, custom-designed and manufactured footwear and/or orthotic appliances and advice on the need and application of medical-grade footwear, orthotic appliances and other footwear.

# 1. Introduction

Diabetes-related foot ulcers are a significant complication that can lead to serious consequences, including lower extremity amputation and even mortality [1]. Diabetic foot ulcers have become a major burden on the healthcare industry, with limited treatment options and a high risk of progressions [2]. Furthermore, diabetic foot ulcers not only reduce the quality of life for patients but also impose a substantial financial burden on individuals and society as a whole [3]. Treatment of diabetic foot ulcers is protracted and intensive, requiring extensive wound care, infection control, and management of underlying vascular disease [4]. This necessitates the need for innovative approaches to prevent and manage diabetic foot ulcers effectively.

One promising approach is the use of AI-driven personalised offloading device prescriptions. AI-driven personalised offloading device prescriptions have the potential to revolutionise the prevention and management of diabetic foot ulcers.

By leveraging artificial intelligence, healthcare providers can analyse large amounts of patient data, including medical history, foot morphology, and gait analysis, to develop personalised offloading device prescriptions. This cutting-edge approach allows for the customisation of offloading devices, such as orthotic inserts or specialised footwear, to match the specific needs of each individual patient. This personalised approach is crucial because not all offloading devices are suitable for every patient. The importance of offloading devices in the management of diabetic foot ulcers has been recognised by national and international guidelines [5]. However, the current utilisation of offloading devices in clinical practice is suboptimal. Research conducted in the UK found that only 5% of patients with diabetic ulcers received a pressure-offloading device [5]. This low utilisation of offloading devices could be attributed to various factors, including a lack of awareness among healthcare providers, inadequate training, and challenges in accessing orthotic services [3]. To address these barriers and improve the utilisation of offloading devices, AI-driven personalised offloading device prescriptions offer several key advantages [6]. Firstly, AI-driven personalised offloading device prescriptions can enhance the accuracy and effectiveness of offloading interventions [7].

By taking into account a patient's individual characteristics and needs, AI algorithms can recommend offloading devices that will effectively redistribute pressure and relieve the abnormal load on the plantar foot surface. This personalised approach increases the likelihood of successful ulcer healing and reduces the risk of complications associated with diabetic foot ulcers.

Furthermore, AI-driven personalised offloading device prescriptions can also address barriers to adherence and compliance with offloading devices [8]. For example, one of the barriers to adherence is postural instability caused by wearing offloading devices, which can lead to falls and accidents while performing daily activities [2]. However, by utilising AI algorithms to analyse gait patterns and biomechanical data, personalised offloading devices can be designed to minimise gait disturbances and improve stability, thereby reducing the risk of falls.

The use of AI-driven personalised offloading device prescriptions can also help overcome the challenges in accessing pedorthic services. Due to the shortage of pedorthic professionals and the limited availability of pedorthotic clinics, many patients face difficulties in accessing offloading devices [9].

AI algorithms can bridge this gap by remotely analysing patient data and prescribing personalised offloading devices that can be manufactured using 3D printing or other advanced fabrication.

## 2. Study background

This study is deduced from four earlier series of individual studies, and they are presented in Fig. 1 with further brief descriptions of the individual studies to provide the contexts.

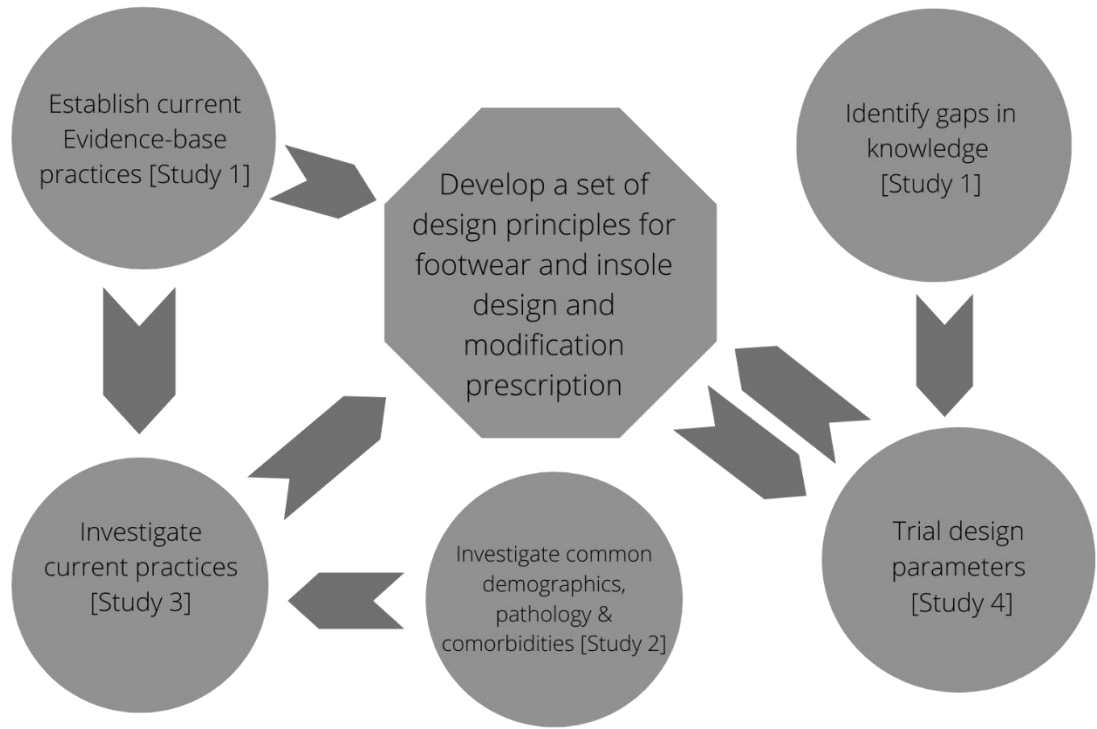

Figure 1. *Schematic of research approach and connections among background studies*

**2.1 Study 1: Systematic literature review**

Twenty-five studies were reviewed systematically [9]. The included articles used repeated measure (n = 12), case-control (n = 3), prospective cohort (n = 2), randomised crossover (n = 1), and randomised controlled trial (RCT) (n = 7) designs. This involved a total of 2063 participants. Eleven studies investigated footwear, and fourteen studies investigated insoles as an intervention. Six studies investigated ulcer recurrence; no study investigated the first occurrence of ulceration. The most commonly examined outcome measures were peak plantar pressure (PPP), pressure-time integral (PTI) and total contact area. Methodological quality varied.

Strong evidence existed for rocker soles to reduce peak plantar pressure. Moderate evidence existed for custom insoles to offload forefoot plantar pressure. There was weak evidence that the insole contact area influenced plantar pressure. Footwear and insoles are complex interventions, and the outcome measure is still limited to PPP reduction and ulcer recurrence. Rocker soles, custom-made insoles with metatarsal additions and a high degree of contact between the insole and foot reduce plantar pressures in a manner that may reduce ulcer occurrence.

Most studies rely on a reduction in PPP measures as an outcome as a proxy for the occurrence of ulceration. There is limited evidence to inform footwear and insole interventions and prescriptions in this population. Further high-quality studies in this field are required. Approaches to measuring patient adherence are lacking but play a vital role in the overall outcome of the treatment.

**2.2 Study 2: Retrospective clinical audit**

A retrospective clinical audit of a cohort of 70 patients at a suburban pedorthics clinic was undertaken to understand the pedorthists' patient profile, including sociodemographic, pathological, comorbidity-related, and other individual characteristics.

The mean age of participants was 64.69 (SD 11.78) years, ranging from 27 to 90 years old. They were more likely to be male (n=43 males (61.4%)). All participants were overweight to obese, with a mean weight (kg) of 91.37 (SD 14.73). The average BMI was 30.96 (SD 4.15).

Most (97.2%) participants had Type-2 diabetes mellitus (T2DM), and only a few (2.8%) had type-1 DM (T1DM). Australia was the birthplace of the highest number of participants (n=28), and the majority of the participants (n=42) were born outside of Australia. About 5.7% (n=4) were of Aboriginal or Torres Strait Islander origin. The mean duration of diabetes among the participants was 14.09 years (SD 6.58). The mean duration of neuropathy was 8.56 (SD 4.16) years.

Approximately 47% (n=33) of participants had HAV; 39% (n=27) participants had hammertoe and cavus foot conditions, and 33% (n=23) of participants had clawed toes. Common foot pathologies among the participants were bony prominence at 71% (n=50), rigid flat foot, and limited joint mobility (LJM) (53%, n=37). Hyperkeratosis was the most common condition in the participant group; everyone (n=70) had this condition. Of previous foot pathology, about half (47%) of the participants had a history of forefoot ulceration. Around one-third, 34% (n=24) of participants, had forefoot amputation, and around 34% (n=24) had undergone a digital amputation.

The most common comorbidities in this group were rheumatoid arthritis (RA) 36%, Peripheral vascular disease (PVD) 41%, lymphodema 20%, and posterior tibialis tendon dysfunction (PTTD) 26%.

The main funding providers for footwear in this population group, comprising 78% (n=55), was Enable NSW, followed by privately (self) funded at 10% (n=7), Closing the Gap at 4.3% (n=3), private health insurance 2.9% (n=2), and aged care package 1.4% (n=1).

This shows the complexity of patients, highlights the variations in social issues, funding models, cultural needs, and personal preferences, and how this might impact the outcome of patient care through appropriate footwear and insoles for their conditions. This guides the variations in the case studies to represent a "typical" male or female patient seen at the pedorthics clinic, particularly the sociodemographic, foot pathology, and comorbidity characteristics.

*Clinical case studies*

The audit results were used to create four 'typical patient' case studies based on the categories of age, gender, country of birth, duration of diabetes and neuropathy, foot pathology, comorbidity, and health fund access provision for representing patients who come to pedorthic clinics for the provision of appropriate footwear and insoles. The cases were verified by an expert panel and incorporated into Study 3 which was used to help understand Pedorthic prescribing practices.

**2.3 Study 3: Australian pedorthist's survey**

The purpose of this study was to examine the current prescription habits of Australian pedorthists when designing and altering footwear and insoles with the goal of offloading for neuropathic plantar forefoot ulcer prevention and improved patient adherence for the four case studies developed in Study 2. The survey questionnaires explored pedorthist's practice in terms of (i) footwear design and modification parameters, and (ii) insole design and modification parameters, including adherence-related challenges for footwear and insoles and their overcoming strategies.

Multiple-choice and open-ended questions were used to explore pedorthist's prescribing behaviour in terms of the case studies. The criteria explored were adopted from Study 1 and Diabetes Feet Australia (DFA) guidelines [10]. Nineteen pedorthists completed the survey (45% of pedorthists).

There was some level of consensus amongst pedorthists around the treatment of patients for the four case studies. The area of highest consistency was the recommendation to use custom-made footwear for case 3, which was a particularly complex case study. The domains that achieved the greatest consensus for treatment amongst pedorthsists in footwear and insole design were upper height and rocker sole design profiles, insole type, and design characteristics. The areas with the least consensus were the prescription of prefabricated medical grade footwear recommendations with or without modifications (cases 1, 2, and 4) and the heel height, toe spring, footwear materials, insole materials, casting methods, and evaluating the pressure offloading efficacy of the devices.

Most of the recommendations by the pedorthists fall within existing accepted practice guidelines, and variations were likely to be due to the different pedorthist training and the scope of practice [11]; available options of footwear type supply in their practices and variation in material supply for manufacture and modifying insoles; health fund availability; patient's preferences; and intended activity.

The findings from this study were also limited by the relatively small sample. This, in part, reflects the small size of the profession, so the overall possible numbers of responses were limited, but a large variation is likely with a small sample size.

This study highlights the complexity of footwear as an intervention for people with diabetes-related foot disease and the high level of variation possible because of the multiple components associated with shoes. Given that the primary goals of footwear as an intervention are to prevent injury (largely by accommodating existing foot deformities) and reduce plantar pressure, it is clear that there are a number of different routes to achieving this goal, which will be impacted heavily by patient preference and adherence. Therefore, a less clinically prescriptive approach may be necessary that takes into account a range of social and more subjective factors, such as patient preference and goals, activity levels, funding source, and availability, and the availability of different materials and footwear – this then leads into the n=1 study that varied a number of attributes of footwear with a goal of achieving the greatest reduction in plantar pressure while optimising patient adherence.

The results present a lot of variation in clinical recommendations for the same patient; however, they appear largely based on valid considerations and assumptions. The variations in recommendations were in footwear type, upper height, heel height, toe spring, rocker sole design profile, insole casting method, insole materials, and insole design parameters. This also highlights the need for an evidence-based guideline to guide practice and help reduce variations in clinical practice or provide guidance that can increase the consistency of prescribing patterns. Evidently, there is no 'one size fits all' – and there is probably no 'rule' to dictate that. The results of this study, together with the results from Study 1 (9), have been used to form the knowledge base for footwear and insole design and modifications and to test those parameters in Study 4 towards recommending a set of design principles for footwear and insole design and modification prescriptions.

**2.4 Study 4: A series of N-of-1 trials**

Building on the learning from the previous studies, which demonstrate, first, the diversity and complexity of patient needs and preferences, and second, the wide range of treatment options to achieve the common goals of optimal protection of the foot and a reduction in plantar pressure, this study [12] used a patient-centred intervention approach and study design, the N-of-1 trial (a series of). This study allowed the application of a range of variables in footwear and insole design tailored to the individual patient's needs, with a view to achieving optimal pressure reduction and adherence.

The series of N-of-1 trials included 12 patients that formed 12 individual N-of-1 studies. Two footwear prototypes and three insole prototypes, each with some level of customisation for the patients, were applied and modified over no more than three iterations.

The interventions were fully custom-made footwear (Shoe A) and prefabricated medical-grade footwear (Shoe B) with modifications. There were three different types of custom-made insoles: Insole A (to follow custom shoe last plantar profile through heat moulding method), Insole B (3D printed insole base from TPU filament combined with Poron and EVA/Plastazote top cover (for Shoe B), Insole C through conventional heat moulded manufacturing method to (fit into shoe B). All the footwear and insoles underwent a series of modifications guided by in-shoe pressure mapping and patient feedback on balance and suitability for purpose. Individual treatment goals (target PPP) were met for each case, and the exploration for variations in enhancing adherence continued.

Barefoot static and dynamic pressure analysis and in-shoe pressure analysis on the baseline footwear were done at the initial appointment (T0). The intervention footwear and insole design were decided at T0. Intervention footwear and insole were fitted at the $2^{nd}$ appointment (T1). A maximum of three rounds of modifications were carried out on the footwear and insoles until an acceptable plantar pressure offloading threshold was achieved. Patient satisfaction and adherence-related information were captured at each appointment (T1-T4). The results show that with tailored responses to individual patient needs, PPP can be reduced substantially, and there is a strong need to consider multiple, complex patient issues to enhance adherence.

It is already proven that there is no panacea for footwear and insole prescription; instead, there are a series of principles based on, first, patient needs and preferences, and second, patient pathology. Those are the guiding factors for the treatment plan and options. These complex factors around patients' pathology, comorbidity, and personal and social perspectives need to be put in the bigger picture, and the design principles proposed in the study have considered all these factors for improved clinical and patient adherence outcomes.

## 3. Design Principles

Here, the design principles for footwear prescription for people with diabetes-related foot disease and at risk of neuropathic plantar forefoot ulceration that arose from the above studies are outlined. These principles underpinning footwear and insole design aim to guide pedorthists involved in prescribing footwear for people with diabetes-related foot disease to prescribe and produce footwear based on the best evidence for plantar pressure offloading and strategies to improve patient satisfaction and adherence. The outcomes of Studies 1 (systematic review) and 3 (Australian pedorthists survey) provided the knowledge base of various footwear and insole design and modification parameters in the literature and in real practices by the pedorthists. The common agreements and the variations were both noted, and patient adherence-related challenges and overcoming strategies were also noted (Study 3). Then, these parameters were tested further and explored the outcome on individual patients through a series of N-of-1 trials to establish more specific design and modification parameters for specific forefoot pathologies in people with diabetes and neuropathy, and their adherence-related factors to improve the outcomes were also established.

These parameters were then presented in a patient-centric Clinical Decision Support Database (CDSD) for footwear and insole prescribing to ensure the prescription is made based on the most suitable option for the individual. These are presented in Table 2. This CDSD theme aims to ensure the maximum possible adherence by the person when all possible factors are considered for the individual associated with their therapy and treatment goals. Then, the information or output from this CDSD is taken into the framework of a set of design principles for the technical prescription to ensure optimum clinical outcomes such as plantar pressure offloading and walking comfort, ease of use of the patient and such. This information is aligned with the workflow presented in Fig. 2. The core information for this set of design principles is based on our systematic literature review (Study 1), DFA guideline [10], Australian pedorthists' survey (Study 3), the series of N-of-trials (Study 4). The main framework of the CDSD is based on the results of Study 4 and the other studies, including Study 3 results that have been used to complement the database. Table 2 presents the various evidence-based parameters for the CDSD model, which can be used for clinical decision-making for the pedorthists. These parameters are extracted from Table 1, systematic literature review (Study 1), DFA guideline [10] clinical audit (Study 2), Australian pedorthists' survey (Study 3), the series of N-of-trials (Study 4) and Zwaferink et al.'s algorithm [13].

The treatment goals underpinning this set of design principles are to: (i) Optimise patient satisfaction and adherence to therapy (by improving walking comfort, ease of use and aesthetics, and also considering the personal circumstances of the patients), (ii) Protect the foot from injury and cause no further injury to the foot, (iii) Reduce peak plantar pressure, and (iv) Optimise balance and mobility.

To achieve this treatment goal requires that the patient will wear the footwear >80% of the time [14]; therefore, a further treatment principle is that the footwear needs to meet the aesthetic and/or social requirements of the patient and /or their main decision maker (e.g., spouse or partner).

The footwear needs to be affordable for the patient [15, 16], and the guidance towards available funding is as important as educating the patient on foot self-care [17]. The inability to afford the cost of therapy and having no access to health funds can limit treatment options in this population group [17].

The footwear needs to be fit for its purpose [18]. To increase adherence by achieving the patient's goals and aesthetic requirements, additional individual factors need to be considered as appropriate such as appropriateness for the climate and cultural and religious beliefs.

A summary of the design principles for footwear and insole design and modifications in the form of an infographic to demonstrate the workflow and relevant measures in each step has been presented in Fig. 2.

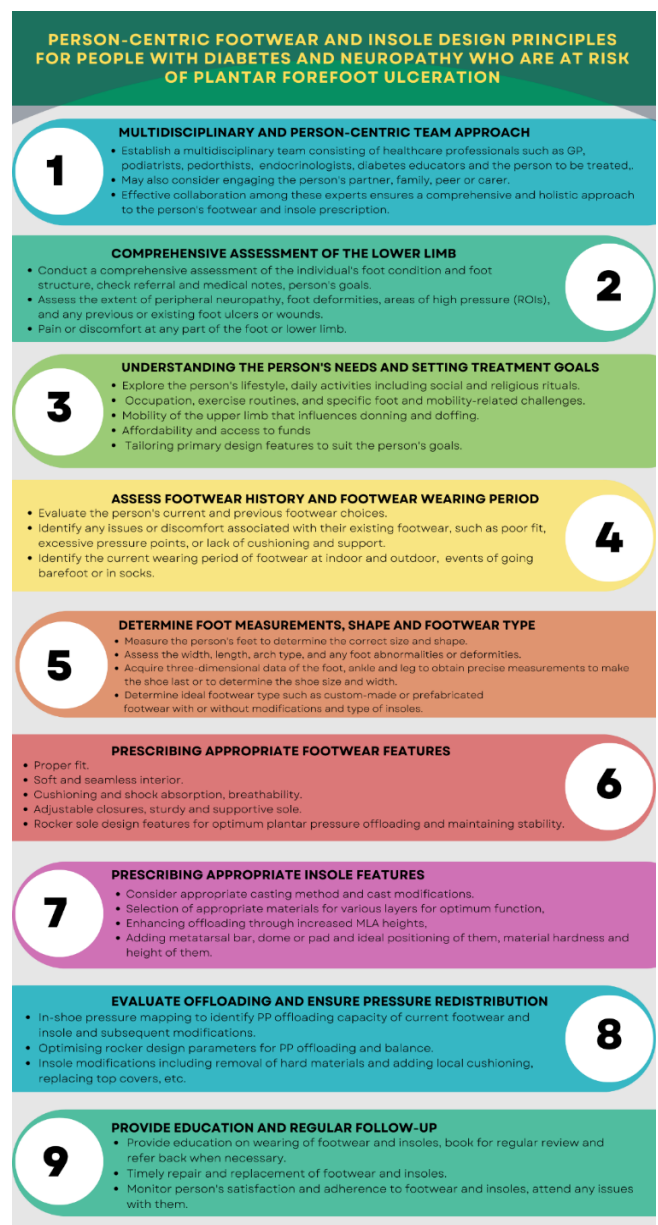

Figure 2. *Workflow and infographic for the design principles for footwear ad insoles*

## 3.1 Multidisciplinary and person-centric team approach

The person to be treated needs to be engaged in the process at the very beginning and needs to be at the centre of the overall activities. All relevant health professionals need to be engaged and provide input into the care plan. Engaging a friend or family or a carer as appropriate is very important for better treatment and adherence-related outcomes. Studies 3 and 4 have demonstrated the evidence of these approaches for a positive outcome. The CDSD model (Table 2) has described the various evidence-based parameters around this step under *"Person's preferences and intended activity"*, *"Person's mobility status"*, and *"Family/partner/carer/peer preferences and advocacy"*.

## 3.2 Comprehensive assessment of the lower limb

A comprehensive assessment of the person's foot condition and foot structure, current mobility status and mobility goals are the second steps in the process. Some of the information is also available in the referral and medical notes. The person's goals recorded in the referral form need to be checked against during the assessment for currency and a clear understanding by all parties involved in the process.

Other assessment aspects involve assessing the extent of peripheral neuropathy, foot deformities, areas of high pressure (ROIs), and any previous or existing foot ulcers or wounds. Attention should be paid to the person reporting on any pain or discomfort at any part of the foot or lower limb or during any specific activities or mobility phases. The CDSD model (Table 2) has described the various evidence-based parameters around this step under *"Main foot pathology", Comorbidity"*, and *"Body weight"*.

## 3.3 Understanding the person's needs and setting treatment goals

Gathering information on the person's lifestyle and daily activities, including social and religious rituals, is critical for a comprehensive treatment plan and device design workflow. Understanding the person's needs relating to occupation, exercise routines, and specific foot and mobility-related challenges are integral parts of the workflow. Assessing the mobility of the upper limb is important to identify the person's ability that influences donning and doffing. A person's affordability to the therapy and access to funds are also important factors to consider when designing the treatment plan. Studies 3 and 4 report on the importance of fund access for these populations. From this step, tailored primary design features to suit the person's goals can be drawn. Studies 3 and 4 have demonstrated evidence of these workflows. The CDSD model (Table 2) has described the various evidence-based parameters around this step under *"Person's preferences and intended activity"*, *"Person's mobility status"*, and *"Family/partner/carer/peer preferences and advocacy"*.

## 3.4 Assess footwear history and footwear wearing period

It is important to encourage the person to bring any current and old footwear to evaluate the footwear choices and gait patterns in real life. A thorough assessment to identify any issues or discomfort associated with their existing footwear, such as poor fit, excessive pressure points, or lack of cushioning and support, needs to be conducted. An in-depth exploration in identify the current wearing period of footwear at indoor and outdoor events of going barefoot or in socks is critical for appropriate device design principles and setting education goals. Studies 3 and 4 have demonstrated evidence of these workflows that influence the treatment plan and device design specifications. This information complements the CDSD parameters under *"Person's preferences and intended activity"*.

### 3.5 Determine foot measurements, shape, and footwear type

Measuring the person's feet by using any suitable methods to determine the correct size and shape is critical for footwear selection. The width, length, arch type, and any foot abnormalities or deformities are the guiding factors for the correct size, width, and type of footwear to be recommended. The measuring process involves acquiring three-dimensional data of the foot, ankle, and leg to obtain precise measurements to make the shoe last or to determine the shoe size and width. This helps determine the ideal footwear type, such as custom-made or prefabricated footwear with or without modifications and the type of insoles required. The ratio of rearfoot volume and forefoot volume is critical to determine the appropriate footwear type to optimally offload the forefoot PPP. For example, a Cavus foot with narrow rearfoot and wide forefoot should be recommended a fully custom-made footwear and insole for a better fit and reduce shear that a prefabricated medical grade footwear in unable to deliver. Very often, the prefabricated medical-grade footwear is too loose at the back when an adequate forefoot width is chosen for this type of foot. Some participants in Study 4 have demonstrated evidence of such needs. This information helps develop and select the CDSD parameters under *"Foot structure and shape",* and *"Footwear type"*.

### 3.6 Prescribing appropriate footwear features

Proper fit of the footwear is the most critical factor in offloading and adherence to the therapy, and the benefits multiply when they are equipped with appropriate design features such as a soft and seamless interior. Other critical design features for the footwear are cushioning and shock absorption, breathability, adjustable closures, and sturdy and supportive soles. Strong evidence is present for rocker sole design features for optimum plantar pressure offloading and maintaining stability, as reported in Studies 1, 3 and 4.

This information helps develop the CDSD parameters under *"Footwear style", "Footwear upper height", "Footwear lining material", "Footwear fastening system", "Footwear upper flexibility", "Footwear tongue flexibility", "Footwear heel counter", "Footwear heel height", "Footwear heel modification", "Footwear outsole", "Footwear rocker sole profile", "Footwear apex position", "Footwear apex angle", "Footwear rocker angle".* The abovementioned parameters are extracted from our systematic literature review (9)*,* DFA guideline (10), clinical audit [Study 2], Australian pedorthists' survey [Study 3], the series of N-of-trials [Study 4] and Zwaferink et al.'s algorithm (13).

### 3.7 Prescribing appropriate insole features

The selection of an appropriate casting method and cast modifications are influential factors in outcomes when designing an insole that is optimal for PPP reduction and increasing contacts.

The selection of appropriate materials for various layers can achieve the above goals. Other strategies for enhancing offloading are through increased MLA heights, adding metatarsal bar, dome or pad and ideal positioning of them, material hardness and height of them. Studies 1 (9), 3 and 4 have confirmed the features and benefits of such strategies.

This information helps develop the CDSD parameters under *"Insole type"*, *"Custom-made insole shape"*, *"Insole base layer material"*, *"Insole mid-layer material"*, *"Insole top layer material"*, *"Insole heel cup"*, *"Insole heel wedge"*, *"Insole MLA height"*, *"Insole metatarsal addition"*, *"Insole metatarsal addition position"*, *"Insole metatarsal addition thickness"*, *"Insole metatarsal addition material type"*, and *"Insole modification"*.

**3.8 Evaluate offloading and ensure pressure redistribution**

It is very important to evaluate the efficacy of PPP offloading and redistribution of the designed devices through in-shoe pressure mapping at the fitting of current footwear and insole and during any subsequent post modifications. Various strategies are successful in increasing the offloading capacity of the footwear, and rocker design parameters for PP offloading and balance are the most popular ones. Its efficacy and design features have been reported in Studies 1, 3 and 4.

Among other popular strategies to increase PP offloading are the insole modifications that include the removal of hard materials, adding local cushioning, and replacing top covers. Studies 1, 3 and 4 have reported the features and benefits of such strategies. This information helps develop the CDSD parameters under *"Pressure offloading evaluation method"*.

**3.9 Provide education and regular follow-up**

The success of the treatment plan and the footwear and insoles are largely dependent on education on wearing footwear and insoles, regular reviews, repair, maintenance, and timely Replacement. It is also important to refer to a podiatrist or other relevant health professional when necessary.

Person's satisfaction and adherence to the devices can vary depending on various factors, and it is critical to keep monitoring a person's satisfaction and adherence to footwear and insoles. If there are any concerns or issues reported by the person or observed during the appointments, it is critical to attend to any issues that arise.

*Foot pathologies associated with neuropathic plantar forefoot ulcers*

The table below describes the common pathology seen in people with diabetes and neuropathy and associated with plantar forefoot ulceration. This is a general guide for the pedorthists for what pathology this algorithm is suitable for. This information is collated from the systematic literature review (Study 1).

**Table 1.** *List of forefoot pathology based on the literature review.*

| Foot Pathology | Description |
|---|---|
| **Neuropathy** | "The neuropathic foot is described as a loss of peripheral nerve function, which can be sensory, motor, autonomic or, usually, a mixture. This loss of function leads to structural changes and function of the foot towards ulceration and subsequent amputation." |
| **Hyperkeratosis** | "It is commonly called calluses, and the formation of calluses is due to repeated excessive pressure on the skin. In patients with neuropathy, the presence of a callus increases peak plantar pressure and increases the risk of ulceration in that area. Calluses are commonly seen in diabetic feet, even in the absence of neuropathy." |
| **Bony prominences at metatarsal heads** | "Claw and hammer toes are associated with plantar fat pad displacement and metatarsal head prolapse on the plantar surface. Any ulcers in the metatarsal heads need to be treated with urgency, especially in the hallux base, due to the increased risk of amputation." |
| **Hallux Abducto Valgus (HAV)** | "Due to the structural deformity caused by HAV and the abnormal foot shape, the normal push-off becomes difficult and results in increased friction on the medial aspect of the 1st MTP Joint." |
| **Flexible flat gait foot** | "Flexible flatfoot results in reducing the shock-absorbing capacity of the foot and increases pressure on the medial border." |
| **Rigid flat foot** | "The rigidity of this condition results in excessive pressure on the medial border of the foot. Ankle-high shoes with shock absorber heels, stronger medial heel counter, and rocker with apex position posterior to metatarsal heads are ideal features to protect the foot from worsening in positioning." |
| **Forefoot amputation** | "There are many similarities in the effect of forefoot amputation with Hallux amputation, with the additional risk of the foot taking an equines structure and increased pressure at the lateral border of the foot. (Sage, Pinzur, Cronin, Preuss, & Osterman, 1989) The shock absorption capacity decreased due to the stiffness of the foot structure." |
| **Hallux amputation** | "Amputation of the Hallux results in altered pressure distribution, and the pattern is significantly influenced by this (Lavery et al., 1995). During the push-off phase, the force is transferred through the 1st metatarsal bone and results in increased shear force. This mechanism frequently results in a wrinkle on the shoe's upper and pressure ulcers on the dorsal aspect of the foot." |
| **Neuropathy** | "The neuropathic foot is described as a loss of peripheral nerve function, which can be sensory, motor, autonomic or, usually, a mixture. This loss of function leads to structural changes and function of the foot towards ulceration and subsequent amputation." |
| **Hammer & clawed toes** | "A typical neuropathic foot with stiff structure and minimal shock absorbing and contact area due to the dorsiflexed position of the Metatarso Phalangeal Joints (MTPJ's)." |
| **Limited joint mobility** | "Limited joint mobility in the diabetic foot has been described by the limited range of motion (ROM) at the ankle joints and 1st Metatarso Phalangeal Joints (MPJ) (Boffeli et al., 2002; Lobmann et al., 2002; Murray et al., 1996; Nube et al., 2006; Van Gils & Roeder, 2002) Ankle joint limited ROM or equinovarus foot structure increases the pressure at the forefoot area, specifically at the metatarsal zone, which accelerates the risk of ulceration in that area. In addition, Hallux limitus or rigidus can generate foot ulcers in the medial and dorsal aspects of the 1st Hallux (Lázaro-Martínez et al., 2014). As the foot is stiff in nature (Delbridge et al., 1988), the force is transferred |

| | through the heel during heel strike yielding less shock absorption within the foot at the gait cycle. As the forefoot has limited dorsiflexion, that results in friction between the forefoot and shoe at the push-off phase." |
|---|---|

## 4. Design considerations

**4.1 Upper design**

For footwear upper height, the following classification is used:

a. Low cut = below the malleolus.
b. High cut = at the level of the malleolus.
c. Extra high cut = above the malleolus and up to the knee.

The purpose of the adequate upper height of footwear is to influence forefoot plantar pressure reduction and accommodation of the feet.

*Principles to prevent or reduce injury*

**Current evidence**: Some common principles need to be considered when prescribing footwear for people with diabetes and neuropathy (13). Each shoe must have sufficient interior space in length and width, with a minimum of 1 cm space in length between the longest toe and the inner of the shoe. The toe box must be sufficiently high to accommodate a non-correctable claw, hammer toes, or a hyperextended hallux. The inner lining should not have any seams. Shoes should have laces or velcros depending on hand functions, and BOA lacing and zippers can be considered for further support to aid foot entry.

*Management* of *oedema*

**Current evidence**: With the presence of oedema or vulnerable skin in the lower leg, a low-cut shoe is preferred to avoid pressure from the shoe upper or the top line on the sensitive area. If a high-cut shoe is indicated, the inner should be padded, and the top edge of the shoe should be above the vulnerable area. A 1.5mm-thick flat layer of material (single or multiple layers as needed) below the insole creates the opportunity to moderate interior volume with changing oedema [13].

**4.2 Footwear upper flexibility**

**Treatment goal:** The treatment goals related to upper flexibility are the accommodation of the feet, supporting feet structure and ensuring walking comfort and ease of use. Patient stability increase, reduced risk of falls, and improved balance are some of the key focuses for footwear upper design.

**Current evidence**: If the lower limb pathology has a combination of muscle weakness of the tibialis anterior or peroneus longus muscles, an extra high cut upper with reinforcement between the upper and lining should be considered [9, 13]. This is to support the dorsiflexion of the ankle.

Another alternative approach is to add an external orthosis or bracing in the form of an ankle-foot orthosis (AFO) when a foot drop is present.

### 4.3 Rocker sole profile

**Treatment rationale:** The purpose of a rocker sole is to reduce peak pressure under the forefoot [9] by redistributing the plantar pressure.

- With fully custom-made shoes, the rocker is applied in the insole, with the outsole following this insole rocker configuration one-on-one.
- With prefabricated medical-grade footwear, the rocker is in the outsole.

**Current evidence for rocker apex position**: The rocker apex is the central point on the rocker axis and should be at 60-65% of the shoe length or 10-15mm behind the metatarsal heads (MTHs) [9]. The % of the length is measured from the rear of the shoe to provide optimal balance for pressure relief under the different metatarsal heads. This relates to a rocker axis that is ~1.3cm proximal to MTH 1 and ~2.6cm proximal to MTH 2 for shoe size US9.5 [13]. Barefoot pressure mapping or a pedograph or in-shoe pressure mapping, or both should be considered for the further precision design of the rocker profile.

*Considerations for severe neuropathy or poor balance*

The rocker apex position must be set carefully for people with severe neuropathy and poor balance. A distal apex location may help support them during the stance phase.

**Current evidence for rocker angle**: The rocker angle is the angle between the ground and the bottom surface of the shoe from the rocker apex forward and should be 15-20° in each shoe, independent of shoe size [13]. This should be guided by pressure mapping or pedograph and checking the person's balance. The thickness of the rocker sole also needs to be considered with the aim of fall risk assessment and aesthetics of the footwear. This should determine the precise rocker angle for the prescribed footwear.

**Current evidence for apex angle**: The apex angle is the angle between the rocker axis and the longitudinal axis of the shoe and should be 95° [9, 19]. This means that the rocker axis is medially more distal than laterally. With an exhortation position of the foot, the rocker axis must be corrected to give an apex angle of 95° in the direction of walking [13]. This angle may change where the metatarsal head orientations are unique, and the ROI for offloading is unique.

### 4.4 The outsole profile

**Treatment rationale**: The purpose of the outsole is to protect the midsole that contains the rocker profile and support the rocker profile structurally. This is the interface between the walking surface and the other parts of the footwear that accommodate the foot. It is also one of the most visible and visual factors affecting patient satisfaction and adherence.

A number of different outsole options are possible. There is not sufficient evidence to be prescriptive about the outsole except the shape of them based on practical requirements, such as with a separate heel or in a wedge shape. A separate heel is considered for the patient's aesthetic preference or occupational needs, and the wedge shape is considered when the base of ground contact needs to be more, and the stability and balance of the patient are the priorities.

The footwear outsole should provide cushioning and can be made supple or toughened or can be reinforced with a carbon, fibreglass or metal layer over the partial or entire length of the footwear to create a rigid outsole profile that cannot be bent. The shoe outsole should have adequate shock absorption characteristics while providing sufficient durability for active users and be as lightweight as practically possible. With clearly reduced proprioception, opt for a semi-rigid outsole for improved balance. It is important to consider slip-resistant outer soles for people with moderate to severe peripheral neuropathy.

### 4.5 Tongue

The footwear tongue can be made supple or reinforced, and it is always padded. A rigid tongue with thermoplastic material reinforcement is used mainly with forefoot amputation.

### 4.6 The heel height of the footwear

**Treatment rationale:** The heel of the footwear can have several configurations and this aid with ankle ROM and stability. Changes in heel height influence forefoot PP and stability of the person.

**Current evidence**: Normal heel height for men is 1.5-2cm, and for women, 2.5-3cm in regular prefabricated footwear [13]. Our findings from Study 4 are to have a heel height between 1-1.5cm for improved offloading at the forefoot. An increased heel lift or height is provided in fully custom-made shoes via heel lift in the shoe and in prefabricated shoes via heel lift in the insole (maximum 1 cm) [13]. This can be limited if the footwear is a low-cut version. The increased heel lift in pes equines is dependent on the available ankle range of motion.

### 4.7 The insole design, material and modification features

**Treatment rationale**: Insole can provide base or surface of foot contact, support the medial and transverse arches and accommodate any bony prominences through an appropriate deflection and combination of multiple cushion materials. They also provide cushion to the overall foot, reduce shock during weight-bearing, help stabilise the foot and reduce shear when objectively designed.

The casting method for capturing the plantar foot profile is recommended to be a non-weight-bearing or semi-weight-bearing cast and 3D scan with the aim of further correction of the cast digitally, where possible, for an improved outcome. This concept is verified by our Study 3. The other casting method that can sometimes be recommended is a full-weight-bearing cast when indicated and can be filled with plaster or 3D scanned to make the mould for insole production as practical for the facilities. Generally, a conventional heat moulding of multilayered and multidensity materials is used for a plaster cast mould. For the 3D scanned and designed process, the output can be either by CNC milling of multi-density and multilayered block or 3D printing method out of soft filament or powder with specific geometric pattern or lattice design.

**Current evidence**: The base of the insole in fully custom-made footwear should be with good structural strength capable of shape retention during manufacturing and providing support during everyday use by the patient. This layer also provides the base layer for the mid-layers and top cover to form the complete insole. The hardness of the base layer can be from 55° Shore A onwards, such as a 5-mm-thick micro cork. The base layer can be of dual density with multilayers, and the upper of the base layer materials hardness can be 35-40° Shore A, such as a 5-mm-thick Ethylene Vinyl Acetate (EVA). This layer provides shock absorption properties during weight-bearing. With prefabricated pedorthic footwear, the base may consist of 6-mm-thick EVA (35-40° Shore A) [13]. Any other suitable materials with similar compressibility and durability can also be considered [20]. A 3D printed three-quarter or full-length base with thermoplastic polyurethane (TPU) filament of 45-55° Shore A or a filament with similar functionality can also be used when a 3D print insole is considered.

The insole mid-layer primarily provides shock absorption during weight-bearing and contouring to the plantar foot profile for increasing base of contact and may be made of a 3-6mm thick Poron or PPT. The hardness of the mid-layer can be between 30-35º Shore A. The thickness is dependent on whether custom-made footwear or prefabricated footwear is considered, and the level of cushioning required for optimum pressure offloading of the specific foot.

The top layer of the insole is recommended to provide cushion and sometimes specifically to reduce shear. This layer can be made out of Plastazote or similar characteristic material, and the thickness can be 3-5mm and is also dependent on footwear type and offloading requirements. A Plastazote is more effective in pressure offloading than the leather insole top cover [21]. Other patient-specific factors may be considered when choosing an appropriate type of top cover materials from the range of commercially available materials.

**4.8 The metatarsal additions (Metatarsal bar, pad or dome)**

**Treatment rationale:** Metatarsal additions can help reduce plantar pressure at the metatarsal area significantly [9].

**Current evidence**: A transmetatarsal bar is recommended to offload all metatarsal heads [22]. A metatarsal bar or dome is recommended if only one metatarsal head is the ROI to offload the plantar pressure [13]. The material of the metatarsal additions should be 5-11mm thick [9], made out of PPT or PORON, TPU (or similar 3D printable filament) with 30-35º Shore A hardness. These configurations are proven to be effective and more comfortable for the persons, as found in our study four. The addition is covered by the insole top cover. The location of the addition should be 6-11 mm proximal to the metatarsal head in a static position [9, 23]. Consider the top cover thickness, as this will change the effective position of the addition, moving it more distally.

**4.9 The medial arch support**

**Treatment rationale:** A medial foot arch support is proven to reduce a greater amount of peak plantar pressure at the forefoot [9].

**Current evidence**: Addition of 3-5mm height to the existing foot medial arch support obtained from the total contact through a semi-weight-bearing cast or scan [21, 24, 25]. Full-length medial arch support, in combination with a full-length varus wedge, can improve plantar pressure offloading under the Hallux [26].

When a plantar fascia is tightly tensioned or nodules in the fascia (e.g., in patients with Morbus Ledderhose (plantar fibromatosis)), customisation of the medial arch support should be considered. Support at the sustentaculum tali could be an alternative approach [13].

## 4.10 The insole modification

**Current evidence:** The cut-out should be circular or slightly oval in shape in the walking direction and be minimally larger than the ROI. The cut-out should be 5mm deep and padded with a 3mm durable material up to $30^0$ Shore A (13). The top cover of the insole should be checked regularly and replaced as needed. The replacement frequency of the top cover could be between three to six months, depending on the use and requirements of positioning or adding the metatarsal additions (9, 21).

**Table 2.** *Person-centric CDSD for footwear and insole prescription for people with diabetes and at risk of neuropathic plantar forefoot ulceration*

| CDSD Parameters | Descriptions |
|---|---|
| **Person's preferences and intended activity (PPIA)** | Low-cut casual shoes for outdoors and walking (PPIA1), A low-cut dress shoe (PPIA2), A low-cut indoor shoe with a soft fabric upper (PPIA3), High-cut casual shoes for outdoor and walking (PPIA4), A high-cut dress shoe (PPIA5), Low-cut summer sandal or shoe (PPIA6), High-cut summer sandal or shoe (PPIA7), Extra high-cut casual shoes for outdoor and walking (PPIA8), An extra high-cut dress shoe (PPIA9), Extra high cut summer shoe or sandal (PPIA10), Extra high cut reinforced upper for drop foot (PPIA11), Separate AFO for drop foot (PPIA12) |
| **Foot structure and shape (FSS)** | Normal (FSS1), Wide (FSS2), Very Wide (FSS3), Narrow heel, wide forefoot (FSS4), Swollen rearfoot, narrow forefoot (FSS5), Mismatch foot shape (FSS6) |
| **Main foot pathology (MFP)** | Limited joint mobility of the ankle (MFP1), Pes cavus and claw toes (MFP2), Claw and hammer toes (MFP3), Flexible pes planus with hallux valgus (MFP4), Rigid pes planus with hallux valgus (MFP5), Hallux Rigidus (MFP6), Hallux Limitus (MFP7), Pes equines (MFP8), Hallux or toe amputation (MFP9), Forefoot amputation (MFP10) |
| **Comorbidity (CM)** | PAD/PVD (CM1), Drop foot (CM2), Lower limb edema (CM3), Higher BMI (CM4), Poor vision (CM5), Renal disease, needing dialysis (CM6), History or at risk of falls (CM7), Leg length discrepancy (CM8) |
| **Person's body weight (PBW)** | 60-75 Kg (PBW1), 76-90 Kg (PBW2), 91-110 Kg (PBW3), 111-130 Kg (PBW4), 131+ Kg (PBW5) |

**Table 2.** *Person-centric CDSD for footwear and insole prescription for people with diabetes and at risk of neuropathic plantar forefoot ulceration (Continued)*

| CDSD Parameters | Descriptions |
|---|---|
| **Person's mobility status (PMS)** | Active at home and indoors (PMS1), Active in the community (PMS2), Mostly staying at home (PMS3), Active outdoors and a regular bushwalker (PMS4), Limited mobility, uses 4WW for balance (PMS5), Limited mobility, uses single walking aid for balance (PMS6), Can reach to the toes easily (PMS7), Both hands and fingers are full functioning (PMS8), Single hand and fingers are full functioning (PMS9) |
| **Family/partner/carer/peer preferences and advocacy (FCPA)** | Family/partner/carer/peer agrees to person's choice (FCPA1), Family/partner/carer/peer does not agree to person's choice being impractical or contradicting and advocates towards practitioner's recommendations (FCPA2), Family/friend/carer agrees to person's choice, but peer does not agree due to impractical or contradicting choices (FCPA3), A common agreement was made following further discussion, motivation and advocacy with all parties on the appropriate footwear choices that person is well accepting (FCPA4) |
| **Fund options (FO)** | Self-fund with the flexibility of pursuing the best recommendations (FO1), Self-fund with limitations or restrictions in pursuing the best recommendations (FO2), Health fund support with a co-payment by the person (FO3), Health fund support without a co-payment by the person (FO4), Non-government organisation (NGO) or donor's support for funding (FO5) |
| **Fund options influence footwear type selection (FOIS)** | Fund options influence the footwear type selection (FOIS1), Fund options do not influence the footwear type selection (FOIS2), Fund options partially influence the footwear type selection (FOIS3) |
| **Footwear type (FWT)** | Fully custom-made (Orthopedic medical-grade footwear) (FWT1), Prefabricated medical-grade footwear (pedorthic footwear) without any further modification (FWT2), Prefabricated medical-grade footwear (pedorthic footwear) with further modification (FWT3) |

**Table 2.** *Person-centric CDSD for footwear and insole prescription for people with diabetes and at risk of neuropathic plantar forefoot ulceration (Continued)*

| CDSD Parameters | Descriptions |
|---|---|
| **Footwear style (FWS)** | Casual shoe (FWS1), A dress shoe (FWS2), Indoor shoe (FWS3), Walking shoe (FWS4), Leisure shoes e.g. Golf, lawn bowling (FWS5) |
| **Footwear upper height (FWUP)** | Low cut (FWUP1), High cut (FWUP2), Extra high cut (FWUP3), Slide (FWUP4) |
| **Footwear lining material (FWL)** | Soft Leather lining (FWL1), Micro-fabric with a padded back (FWL2), Mesh with a padded back (FWL3) |
| **Footwear fastening system (FFS)** | Lace (FFS1), Velcro (FFS2), BOA lacing (FFS3), Lace or velcro with a medial zipper for easy foot entry (FFS4), Lace or Velcro with lateral zipper for easy foot entry (FFS5), Lace or Velcro with medial and lateral zippers for easy foot entry (FFS6), Hook & Dow Stick with Velcro (FFS7), Hook & Dow Stick with zippers and larger ring with the runner (FFS8) |
| **Pressure offloading evaluation method (POEM)** | In-shoe pressure analysis (POEM1), Clinical experience and observations (POEM2), Ulcer recurrence (POEM3) |
| **Footwear upper flexibility (FWUFL)** | Suppled (FWUFL1), Rigid (FWUFL2), Stiffened/Reinforced (FWUFL3) |

**Table 2.** *Person-centric CDSD for footwear and insole prescription for people with diabetes and at risk of neuropathic plantar forefoot ulceration (Continued)*

| CDSD Parameters | Descriptions |
|---|---|
| **Footwear upper stiffened location (FWUSL)** | Medial (FWUSL1), Lateral (FWUSL2), Medial + Lateral (FWUSL3), Not required (FWUSL4) |
| **Footwear tongue flexibility (FWTFL)** | Suppled (FWTFL1), Stiffened/Reinforced (FWTFL2), Standard as comes with footwear (FWTFL3) |
| **Footwear heel counter (FWHC)** | Standard (FWHC1), Medial extended and reinforced (FWHC2), Lateral extended and reinforced (FWHC3), Medial + Lateral extended and reinforced (FWHC4), |
| **Footwear heel height (FWHH)** | Standard (FWHH1), Lowered (FWHH2), Increased (FWHH3) |
| **Footwear heel modification (FWHM)** | Heel rounded (FWHM1), Heel flared (FWHM2), Not required (FWHM3) |
| **Footwear outsole (FWOS)** | Suppled (FWOS1), Semi-rigid (FWOS2), Stiffened/Reinforced (FWOS3) |

**Table 2.** *Person-centric CDSD for footwear and insole prescription for people with diabetes and at risk of neuropathic plantar forefoot ulceration (Continued)*

| CDSD Parameters | Descriptions |
| --- | --- |
| **Footwear rocker profile (FWRP)** | No additional rocker is to be added (FWRP1). Additional rocker profile to be added (FWRP2) |
| **Footwear rocker apex position (FWRAP)** | Rocker apex position standard/just behind the metatarsal heads (FWRAP1), Rocker apex position early/posterior to metatarsal heads (FWRAP2), Rocker apex position delayed/anterior to metatarsal heads (FWRAP3) |
| **Footwear rocker apex angle (FWRAA)** | Standard (FWRAA1), Medial direction (FWRAA2), Lateral direction (FWRAA3) |
| **Footwear rocker angle (FWRANG)** | Standard/12-15° (FWRANG1), Moderate/≥20° (FWRANG2), Severe /≥30° (FWRANG3) |
| **Insole type (INST)** | Prefabricated/Standard insole that comes with the footwear (INST1), Custom-made insole (INST2) |
| **Custom-made insole shape (CMINS)** | Regular shape (CMINS1), Medial wall extended (CMINS2), Lateral wall extended (CMINS3), Medial + Lateral wall extended (CMINS4), Toe modelling for partial (toe/s) amputation (CMINS5), Forefoot modelling for forefoot amputation (CMINS6) |

**Table 2.** *Person-centric CDSD for footwear and insole prescription for people with diabetes and at risk of neuropathic plantar forefoot ulceration (Continued)*

| CDSD Parameters | Descriptions |
|---|---|
| **Insole base layer material (INSBLM)** | Hard/firm (INSBLM1), Medium hard (INSBLM2), Soft (INSBLM3) |
| **Insole mid-layer material (INSMLM)** | Medium soft (INSMLM1), Soft (INSMLM2) |
| **Insole top layer material (INSTLM)** | Soft (INSTLM1), Medium soft (INSTLM2), Very soft (INSTLM3) |
| **Insole heel cup (INSHC)** | Regular (INSHC1), Lowered (INSHC2), Increased (INSHC3) |
| **Insole heel wedge (INSHW)** | Medial (INSHW1), Lateral (INSHW2) |
| **Insole MLA height (INSMLAH)** | As per cast/scan (INSMLAH1), increased (INSMLAH2) |

**Table 2.** *Person-centric CDSD for footwear and insole prescription for people with diabetes and at risk of neuropathic plantar forefoot ulceration (Continued)*

| CDSD Parameters | Descriptions |
|---|---|
| **Insole metatarsal addition (INSMA)** | No metatarsal addition (INSMA1), Metatarsal bar (INSMA2), Metatarsal pad (INSMA3), Metatarsal dome (INSMA4), Morton's extension (INSMA5), Reverse Morton's extension (INSMA6) |
| **Insole metatarsal addition position (INSMAP)** | Standard/just behind the metatarsal heads (INSMAP1), Early/posterior to metatarsal heads (INSMAP2), Standard position for Morton's extension (INSMAP3), Standard position for Reverse Morton's extension (INSMAP4) |
| **Insole metatarsal addition thickness (INSMATH)** | Standard/just supporting the metatarsal heads (INSMATH1), Increased/correcting the metatarsal heads alignment (INSMATH2) |
| **Insole metatarsal addition material type (INSMAMAT)** | Hard/firm (INSMAMAT1), Medium soft (INSMAMAT2), Soft (INSMAMAT3) |
| **Insole modification (INSMOD)** | Removal of hard materials (INSMOD1), Local cushioning (INSMOD2), Replacement of top cover (INSMOD3), No further modification required (INSMOD4) |

The above information in Table 2 has the potential to put through a decision tree through an artificial intelligence (AI) powered database to use for machine learning and to develop an AI-powered clinical decision support system (CDSS) on specific footwear and insole type selection for each person based on their main pathology, comorbidity, preferences, and mobility status (treatment goals),

**Table 3.** *Workflow and output of the CDSD model with a real case scenario*

| | | CDSD Parameters description | Participant 1 | Participant 2 | Participant 3 |
|---|---|---|---|---|---|
| **Decision input** | Person-centric data | Person's preferences and intended activity (PPIA) | PPIA10 | PPIA5 | PPIA4 |
| | | Foot structure and shape (FSS) | FSS2 | FSS4 | FSS3 |
| | | Person's mobility status (PMS) | PMS6 | PMS2 | PMS2 |
| | | Family/partner/carer/ peer preferences and advocacy (FCPA) | FCPA4 | FCPA4 | FCPA1 |
| | Diagnosis related data | Main foot pathology (MFP) | MFP5 | MFP1 | MFP6 |
| | | Comorbidity (CM) | CM1 | CM7 | CM4 |
| | | Person's body weight (PBW) | PBW4 | PBW2 | PBW5 |
| | Fund data | Fund options (FO) | FO3 | FO3 | FO2 |
| | | Fund options influence footwear type selection (FOIS) | FOIS1 | FOIS1 | FOIS3 |
| **Decision output** | Footwear design and modification features | Footwear type (FWT) | FWT 3 | FWT3 | FWT3 |
| | | Footwear style (FWS) | FWS1 | FWS2 | FWS2 |
| | | Footwear upper height (FWUP) | FWUP3 | FWUP2 | FWUP2 |
| | | Footwear lining material (FWL) | FWL2 | FWL3 | FWL3 |
| | | Footwear fastening system (FFS) | FFS4 | FFS1 | FFS1 |
| | | Footwear upper flexibility (FWUFL) | FWUFL1 | FWUPFL1 | FWUPFL1 |
| | | Footwear upper stiffened location (FWUSL) | FWUSL4 | FWUSL2 | FWUSL1 |
| | | Footwear tongue flexibility (FWTFL) | FWTFL3 | FWTFL3 | FWTFL3 |
| | | Footwear heel counter (FWHC) | FWHC2 | FWHC1 | FWHC1 |
| | | Footwear heel height (FWHH) | FWHH2 | FWHH1 | FWHH2 |
| | | Footwear heel modification (FWHM) | FWHM3 | FWHM2 | FWHM2 |
| | | Footwear outsole (FWOS) | FWOS3 | FWOS3 | FWOS3 |
| | | Footwear rocker profile (FWRP) | FWRP2 | FWRP2 | FWRP2 |
| | | Footwear rocker apex position (FWRAP) | FWRAP2 | FWRAP2 | FWRAP3 |
| | | Footwear rocker apex angle (FWRAA) | FWRAA1 | FWRAA2 | FWRAA3 |
| | | Footwear rocker angle (FWRANG) | FWRANG1 | FWRANG2 | FWRANG2 |

**Table 3.** *Workflow and output of the CDSD model with a real case scenario (Continued)*

| | | CDSD Parameters description | Participant 1 | Participant 2 | Participant 3 |
|---|---|---|---|---|---|
| **Decision output** | Insole design and modification features | Insole type (INST) | INST2 | INST2 | INST2 |
| | | Custom-made insole shape (CMINS) | CMINS2 | CMINS1 | CMINS1 |
| | | Insole base layer material (INSBLM) | INSBLM1 | INSBLM1 | INSBLM1 |
| | | Insole mi-layer material (INSMLM) | INSMLM1 | INSMLM2 | INSMLM1 |
| | | Insole top-layer material (INSTLM) | INSTLM1 | INSTLM1 | INSTLM2 |
| | | Insole Heel cup (INSHC) | INSHC1 | INSHC1 | INSHC1 |
| | | Insole Heel Wedge (INSHW) | INSHW1 | INSHW2 | INSHW1 |
| | | Insole MLA Height (INSMLAH) | INSMLAH1 | INSMLAH2 | INSMLAH2 |
| | | Insole Metatarsal addition (INSMA) | INSMA3 | INSMA3 | INSMA5 |
| | | Insole metatarsal addition position (INSMHAP) | INSMAP2 | INSMAP1 | INSMAP3 |
| | | Insole Metatarsal addition thickness (INSMATH) | INSMATH1 | INSMATH2 | INSMATH2 |
| | | Insole metatarsal addition material type (INSMAMAT) | INSMAT3 | INSMAT3 | INSMAT2 |
| | | Insole modification (INSMOD) | INSMOD2 | INSMOD1, INSMOD3 | INSMOD1, INSMOD2 |
| | Pressure Offloading Evaluation | Pressure offloading evaluation method (POEM) | POEM1 | POEM1 | POEM1 |

## 5. Towards AI-powered Clinical Decision Support System

In recent years, the integration of machine learning techniques into healthcare has shown great promise in enhancing clinical decision support systems (CDSS) [27]. This section explores the potential of machine learning techniques in developing CDSS for diabetic foot care. We consider a patient scenario, followed by a detailed analysis and the deduction of two exemplary rules that highlight the capabilities of these advanced systems.

Machine learning, a subset of artificial intelligence, can offer a transformative approach to diabetic foot care [28, 29]. It can empower CDSS to analyse vast datasets, identify intricate patterns, and make data-driven recommendations tailored to each patient's unique needs. Decision tree [30, 31] and random forest [29] algorithms are particularly well-suited for this task, as they excel in decision-making and classification tasks. Decision tree algorithms construct hierarchical trees of decisions based on input features, partitioning data into subsets until a prediction or recommendation is reached. Random forest, an ensemble method, leverages multiple decision trees to enhance accuracy and robustness. These techniques can be applied to various aspects of diabetic foot care, including patient-specific factors, clinical diagnoses, and funding considerations.

Let consider a patient (participant 1) scenario: *A low-cut sandal design was the patient's initial desire, but the clinical requirements suggest extra high-cut sandal design shoes (due to lower limb edema). The family was involved, and a common agreement was made following further discussion, motivation, and advocacy with all parties on the appropriate footwear choices that person was well accepting. The patient had a wide foot structure, limited mobility, and used single hand walking aid for maintaining balance. The patient had rigid pes planus and hallux valgus on the right foot, hallux limitus on the left foot. The patient also had PAD/PVD, lower limb edema and a higher BMI (body weight 120 kg). Patient's affordability was dependent on health fund availability and had access to health funds. A prefabricated medical-grade casual design extra high-cut upper design footwear with further sole modification was planned. Microfabric upper lining suitable for PVD/PAD and edema and velcro fastening systems with medial zipper were selected. The footwear needed a rigid forefoot rocker and a higher-density rigid outsole to withstand the higher body weight. The overall thickness of the sole needed to be lower to reduce weight and improve balance, which was achieved by adding a standard rocker angle and positioning the apex posterior to the metatarsal heads and in the medial rocker direction. The insole was a custom-made insole with a regular shape, a medial extended wall with a lowered heel cup and a medial heel wedge due to edema and pes planus feet. A Bilateral Morton's extensions were added with standard thickness with firm material where the insoles had a hard base, medium soft mid-layer and soft top cover. No additional modifications to the insoles were required. An in-shoe plantar pressure measurement system was used to evaluate the PP reduction efficacy of the footwear and insoles.*

This patient scenario vividly demonstrates the complexity of diabetic foot care and the need for a comprehensive CDSS. The patient's initial preference for low-cut sandals clashed with clinical requirements, highlighting the importance of data-driven decision-making. Family involvement and advocacy further emphasised the need for a holistic approach, integrating both clinical expertise and patient preferences. The patient's unique clinical profile, including wide foot structure, limited mobility, and various foot pathologies, underscored the necessity for personalised recommendations. Machine learning algorithms can analyse such profiles, considering factors like patient mobility, body weight, and comorbidities, to tailor offloading device prescriptions.

Table 3 presents the representation of scenarios of three patients/participants in the clinical decision support database. Below are two exemplary rules that can be deduced from data presented in the table.

Rule 1: Family Matters in Footwear Choices

*If (FCPA equals FCPA4 or FCPA3) and (FOIS equals FOIS3 or (FO equals FO2 or FO3)), then FWT should be FWT3.*

Rule 1 emphasises the importance of family involvement when deciding on the right shoes for diabetic foot care. If the patient's family or loved ones are actively engaged in the decision-making process, and there are certain financial limitations or restrictions, then the CDSS recommends choosing a specific type of footwear that's adaptable to the patient's needs (FWT3). This rule highlights that involving the patient's family and working together to make the best choice for the patient's health is crucial, especially when there are budget constraints or differences in opinion.

Rule 2: Customising Insoles for Complex Foot Conditions

*If (MFP equals MFP3 or MFP4 or MFP5 and FCPA equals FCPA1 or FCPA4, and FOIS equals FOIS1 or FOIS3), then INST should be INST2.*

Rule 2 focuses on creating personalised insoles for patients with complex foot conditions. If a patient has specific foot problems like rigid arches, bunions, or limited movement in the big toe and their family supports their choices, then the CDSS suggests using custom-made insoles (INST2). These insoles are designed to fit the patient's unique foot shape and conditions. This rule underscores the importance of tailoring foot support to the individual's foot issues, especially when their family agrees with their decisions.

These exemplary rules exemplify how machine learning algorithms can deduce recommendations based on a patient's clinical and personal factors, family involvement, and funding options. By analysing historical data and considering the interplay of these variables, the CDSS can provide informed, data-driven guidance for diabetic foot care, optimising patient outcomes and reducing the risk of complications.

Table 3 highlights the complexity of the clinical decision support database for diabetic foot care, which includes nine input features and 21 output features, with each feature capable of having between 2 to 10 different possible values. Developing a CDSS utilising machine learning to handle such a wide range of input features and decision output attributes poses significant challenges. These encompass managing high-dimensional data, ensuring data accuracy and reliability, addressing feature selection and dimensionality reduction, dealing with model complexity and potential overfitting, handling imbalanced data classes, modelling intricate feature interactions, providing explainable predictions, adhering to strict data privacy and ethical standards, integrating the CDSS into clinical workflows, accommodating continuous learning and updates in medical knowledge, and maintaining real-world usability and trust among healthcare professionals. Meeting these challenges necessitates a multidisciplinary approach, close collaboration between machine learning experts and healthcare practitioners, and ongoing monitoring and validation in clinical settings.

## 6. Limitations of the Study

Pedorthics is a small profession, but it plays a vital role in managing long-term plantar pressure offloading for patients with high-risk feet. Various National guidelines and standards [9, 10, 15, 16] have recognised the importance of engaging pedorthists in the multidisciplinary team to bridge the gap and enhance patient care. However, the relatively small numbers of registered and certified pedorthists mean that only small sample sizes were available to understand the maximum variations in prescription and practice habits.

Footwear is a complex intervention that needs to meet clinical and patient personal goals and aesthetics. There are some additional variations that may play roles in the decision-making, such as family or spouse's preferences, health fund availability, climate, and cultural influences. They methods used, and the set of design principles derived from this chapter have set the cornerstone for future studies for various patient groups to explore future findings towards evidence-based guidelines.

The studies in the chapters were conducted during the COVID-19 pandemic, which impacted each study, particularly the number of study participants and the timeliness of this research.

The set of design principles for footwear and insole design and modifications derived from this chapter was set out to try to determine the 'science' of orthopedic footwear manufacture for people with diabetes, but it is not a sole science; it is heavily contextually dependent on social issues and patient preferences. Future research based on this set of design principles can increase the scope of practice for various populations.

## 7. Conclusion

The most recent guideline on footwear and insole design and modification for people with diabetes and neuropathy by Zwaferink et al. [13] is aimed at recommending for up to 80% of the population seen in the clinical environment. This guideline [13] is for fully custom-made footwear only and is only feasible for people in developed countries who have different health education and healthcare systems with a variety of fund options. This is a foot pathology-driven guideline and does not include comorbidity, participant mobility concerns, and preferences on footwear choices [13]. However, there is a 20% gap that those people are likely to have more complex conditions and requirements without a proper personalised guideline [9]. Moreover, all the current guidelines are developed based on the environment of the developed countries with better healthcare systems [10, 13]. A set of design principles that are universally applicable that include the provision for people from different climates and developing countries are non-existent [9]. Hence, our set of design principles is universal and bridges the gap in practice to help the practitioners enable practical decision-making to design personalised footwear and insole for people at moderate to high risk of plantar forefoot ulceration. This set of design principles also includes design and modification features for fully custom-made and prefabricated medical-grade footwear (Pedorthic footwear) provision with further modification to match the affordability, intended activities and increased adherence.

The design principles and knowledge gained from this thesis would benefit future researchers exploring personalised medical device design for other healthcare domains. Further research is encouraged for improved clinical and adherence-related outcomes.

It is already proven that there is no panacea for footwear and insole prescription; instead, there are a series of principles based on, first, patient needs and preferences, and second, patient pathology. Those are the guiding factors for the treatment plan and options. These complex factors around patients' pathology, comorbidity, and personal and social perspectives need to be put in the bigger picture, and the design principles proposed in the study have considered all these factors for improved clinical and patient adherence outcomes.

The proposed AI-driven prescription parameters can bridge the gap in current practice and offer more comprehensive design principles leading to improved prescription for personalised device design for the specific patient group.

# Funding

There was no dedicated funding for this study.

# Conflict of Interest